\documentstyle[aps,prl]{revtex}

\input psfig

\begin{document}
\draft
\preprint{DART-HEP-97/07}
\title{How to count kinks: From the continuum to the lattice and back}
\author{Marcelo Gleiser and Hans-Reinhard M\"uller}
\address{Department of Physics and Astronomy, Dartmouth College,
Hanover, NH 03755, USA}
\date {DART-HEP-97/07~~\today}

\maketitle

\begin{abstract}
We investigate the matching between (1+1)-dimensional nonlinear field theories
coupled to an external stochastic environment and their lattice simulations.
In particular, we focus on how to obtain numerical results which are
lattice-spacing independent, and on how to extract the correct effective
potential which emerges from the simulations. As an application, we study the
thermal production of kink-antikink pairs, obtaining a number density of pairs
which is lattice-spacing independent and the effective barrier for pair
production, {\it i.e.}, the effective kink mass.
\end{abstract}


\vspace{1.cm}

The possibility that the Universe underwent a series of symmetry breaking
phase transitions during the earliest stages of its evolution has 
triggered a great deal
of interest in the application of nonequilibrium statistical mechanics to
cosmology. Of particular interest is the potential role that coherent field
configurations, which arise from the interplay between nonlinearities and
out-of-equilibrium conditions, could have played in shaping the earlier
evolution and present-day structure of the Universe. Examples range from the
nucleation of bubbles in the context of inflation and electroweak
baryogenesis to the formation of
topological defects\cite{KOLB-TURNER}.

Given the relevance of the topic, and the obvious difficulties in performing
experiments in a cosmological context, attempts to investigate the emergence
of coherent field structures rely heavily on numerical 
simulations and possible analogies
with condensed matter experiments\cite{ZUREK-TANMAY}. Here we would like to
focus on the former, namely, on numerical simulations designed to investigate
the emergence of coherent structures in thermal field theories. An obvious
limitation of such an approach is that, although field theories are continuous
and usually formulated in an infinite volume, lattice simulations are discrete
and finite, imposing both a maximum (``size of the box'' $L$) and a minimum 
(lattice spacing $\delta x$) wavelength that can be probed by the simulation.
When the  system is coupled to an external thermal (or quantum) bath,
fluctuations will be constrained within the allowed window of wavelengths,
leading to discrepancies between the continuum formulation of the theory and
its lattice simulations; the results will be dependent on the
choice of lattice spacing.

Parisi suggested that if proper
counterterms were used, this depedence on lattice spacing could be 
attenuated\cite{PARISI}. 
This technique was implemented in a study of 2-dimensional
nucleation by Alford and Gleiser\cite{ALFORD-GLEISER}. However, these studies
still left open the question of how to match the lattice results to the
correct continuum field theory. This is a crucial step if we want to test
numerically certain predictions from field theories of relevance not only for
cosmology but also for condensed matter physics, 
such as decay rates of metastable states
and the production of topological defects. Recently, Borrill and Gleiser (BG)
have examined this question within the context of 2-dimensional critical 
phenomena\cite{BORRILL-GLEISER}. 
They have computed the counterterms needed to render the
simulations indepedent of lattice spacing and have obtained a match between
the simulations and the continuum field theory, valid within the one-loop
approximation used in their approach. Inspired by their results, we decided
to investigate the validity of this method within 
the context of topological defects.
The results presented here 
should be relevant to numerical studies of the formation of topological
defects and their comparison with experiments\cite{LAGUNA-ZUREK},
as well as to elucidating the general nature of the effective
potential which emerges from coupling nonlinear field theories to a
stochastic thermal (or quantum) background.

\section{The method}

The Hamiltonian for a classical scalar field with 
potential $V_0 (\phi)$
and an environmental temperature $T$ is, (with $k_B = c = 1$)
\begin{equation}\label{e:hamiltonian}
{{H[\phi]}\over {T}}={1\over {T}}\int dx\left [{1\over 2}
\left( {\partial\phi\over\partial t}\right) ^2 + {1\over 2}
\left( {\partial\phi\over\partial x}\right) ^2 +
V_0(\phi)\right ]\, .
\end{equation}

Even though 1-dimensional field theories are 
free of ultra-violet divergences, the
ultra-violet cutoff imposed by the lattice spacing will generate a
{\it finite} contribution to the effective potential which must 
be taken into account if we are
to obtain a proper match between the theory formulated in
Eq.\ (\ref{e:hamiltonian}) and its numerical simulation on a
discrete lattice. If neglected, this contribution will compromise the
measurement of physical quantities such as the density of kink-antikink pairs
or the effective kink mass. However, before investigating the particular
example of kink-antikink production, we present the method in its most
general form.

For classical, 1-dimensional finite-temperature field theories,
the one-loop corrected effective potential
is given by the momentum integral \cite{PARISI}
\begin{equation}\label{e.oneloopdef}
V_{\rm 1L}(\phi) = V_0(\phi) + {T\over 2}\int_0^{\infty}{{dk}\over
{2\pi}}{\rm ln}\left [1 + {V_0''(\phi)\over k^2} \right ]
= V_0(\phi) + {T\over 4}\sqrt{V_0''(\phi)} \; .
\end{equation}

As mentioned before, the lattice spacing $\delta x$ 
and the lattice size $L$
introduce long and short momentum cutoffs
$\Lambda = \pi / \delta x$ and $k_{\rm min} = 2 \pi / L$, respectively.
Lattice simulations are characterized by one dimensionless parameter,
the number of degrees of freedom $N = L/\delta x$.
For sufficiently large $L$ one can neglect the effect of $k_{\rm min}$
and integrate from $0$ to $\Lambda$.
For $V_0''\ll\Lambda^2$ (satisfied for sufficiently large $\Lambda$),
the result can be expanded into
\begin{equation}\label{e.1Lseries}
V_{\rm 1L}(\phi,\Lambda) = V_0 + {T\over 4}\sqrt{V_0''} -
{T\over 4\pi}  {V_0''\over \Lambda} +
\Lambda T\,\, {\cal O}\left( V_0''^2\over \Lambda^4 \right) \, .
\end{equation}

As is to be expected for a 1-dimensional system,
the limit $\Lambda \rightarrow \infty$ exists and is well-behaved;
there is no need for renormalization of ultra-violet divergences.
However, the effective one-loop potential
is lattice-spacing dependent through the explicit appearance
of $\Lambda$, and so are the corresponding numerical
simulations.
In order to remove this dependence on $\delta x$, we
follow the renormalization procedure given by
BG \cite{BORRILL-GLEISER};
it is irrelevant if the
$\Lambda$-dependent terms are ultra-violet 
finite ($d=1$) or infinite ($d\geq 2$). 
In the lattice formulation of the theory, we
add a (finite) counterterm to the tree-level potential
$V_0$ to  remove the lattice-spacing dependence of the results, 
\begin{equation}\label{e.Vct.gen}
V_{\rm ct}(\phi)=
{T\over 4\pi}  {V_0''(\phi)\over \Lambda}\; .
\end{equation}
There is an additional, $\Lambda$-independent, counterterm which
was set to zero by an appropriate choice of renormalization scale.
The lattice simulation then uses the corrected potential
\begin{equation}\label{e.Vlatt.gen}
V_{\rm latt}(\phi)=V_0(\phi) +
{T \delta x\over 4\pi^2} V_0''(\phi)\; .
\end{equation}
As we will show later in the context of kink-antikink 
pair production, this lattice
formulation simulates the continuum limit to one 
loop as given by Eq.\ (\ref{e.oneloopdef}).

Note that the above treatment yields two novel results. First, that
the use of $V_{\rm latt}$ instead of $V_0$ gets rid of the dependence
of simulations on lattice spacing. [Of course, as $\delta x\rightarrow 0$,
$V_{\rm latt}\rightarrow V_0$. However,
this limit is often not computationally efficient.] Previous works 
\cite{TRULLINGER}, have explored the
influence of a counterterm quadratic on lattice spacing. However, we note that
for small enough $\delta x$, the limit of interest here, our linear correction is
dominant.
Second, that the effective interactions
that are simulated must be compared to 
the one-loop corrected potential 
$V_{\rm 1L}(\phi)$ of Eq.\ (\ref{e.oneloopdef}); once the
lattice formulation is made independent of lattice 
spacing by the addition of the proper
counterterm(s), it simulates, 
within its domain of validity, the thermally corrected
one-loop effective potential.

\section{Application: Thermal nucleation of kink-antikink pairs}

As an application of the method discussed above
we consider the symmetric double-well potential
$V_0(\phi) = {\lambda\over  4} \left( \phi^2 - \phi_0^2 \right) ^2$.
The excitations of the associated quantum theory have a mass
$m = \hbar \omega = \hbar \sqrt{2\lambda }\, \phi_0$.
Thus, in order for the system to remain in the classical regime, 
the condition $T \gg
\hbar
\sqrt{2\lambda }\,
\phi_0$ must hold.
This constrains the dimensionless temperature $\Theta =
T/(\sqrt{\lambda} \phi_0^3)$ to be larger than $\sqrt{2}\hbar /\phi_0^2$.
For $\Theta \ll \tilde M_k \equiv \sqrt{8/9}$, 
where $\tilde M_k$ is the dimensionless
kink mass corresponding to the tree-level potential $V_0$
\cite{Instantons}, we can expect to have only a dilute gas of kink-antikinks
at thermal equilibrium. With these two conditions jointly satisfied, the
system will also obey $M_K \equiv \sqrt{\lambda}\phi_0^3 \tilde M_k \gg m$,
indicating weak coupling.

The corrected lattice potential is
\begin{equation}\label{e.VLatt}
V_{\rm latt}(\phi) =
V_0(\phi) + {3\over 4 \pi^2} \lambda T \delta x \phi^2\; ;
\end{equation}
simulations using $V_{\rm latt}$ will, in principle, match the continuum theory
\begin{equation}\label{e.V1L}
V_{\rm 1L}(\phi) =
V_0(\phi) + {T \sqrt{\lambda}\over 4} \sqrt{3\phi^2 - \phi_0^2}\; ,
\end{equation}
which has (shifted) minima at
$\pm\phi_{\rm min}(T)$, with $\phi_{\rm min}(T) < \phi_0$.

For the numerical simulations we introduce the dimensionless variables
$\tilde t = \sqrt{\lambda}\phi_0\, t$,
$\tilde x = \sqrt{\lambda}\phi_0\, x$,
and $\tilde\phi = \phi/\phi_0$.
To keep the notation simple we will subsequently suppress the tilde.
The field is prepared as $\phi(t=0) = -1$, and evolved in time
according to a Langevin equation with white noise that incorporates the
environmental temperature $T$ through the fluctuation-dissipation theorem.
The details of this and of the numerical implementation are laid out in
\cite{BORRILL-GLEISER}. The viscosity coefficient $\eta$ has been set to
unity throughout this study. The time step is $\delta t = 0.05$, and
$L = 2100$.
The heat bath takes a time $\Delta t \approx 3$ to achieve
equipartition so that the energy per degree of freedom is $E/N=T/2$.

\section{Results}


{\bf Ensemble average of field.}
For sufficiently low temperatures the simulated field will remain
in the vicinity of the minimum $\phi = -\phi_{\rm min}(T)$ 
for a very long time (compared to typical fluctuation time-scales), until
large-amplitude fluctuations drive portions of the space over the barrier at
$\phi = 0$ and beyond. The subsequent evolution is then the
formation of the first kink-antikink pair. True thermal equilibrium
consists of reaching the final equilibrium kink-antikink density together
with zero mean field. In a lose sense, this situation corresponds to symmetry
restoration, although in one spatial dimension ``symmetry restoration''
will occur for any nonzero temperature; it is all a matter of time.

\begin{figure}[c]
\begin{minipage}[b]{.46\linewidth}
\psfig{figure=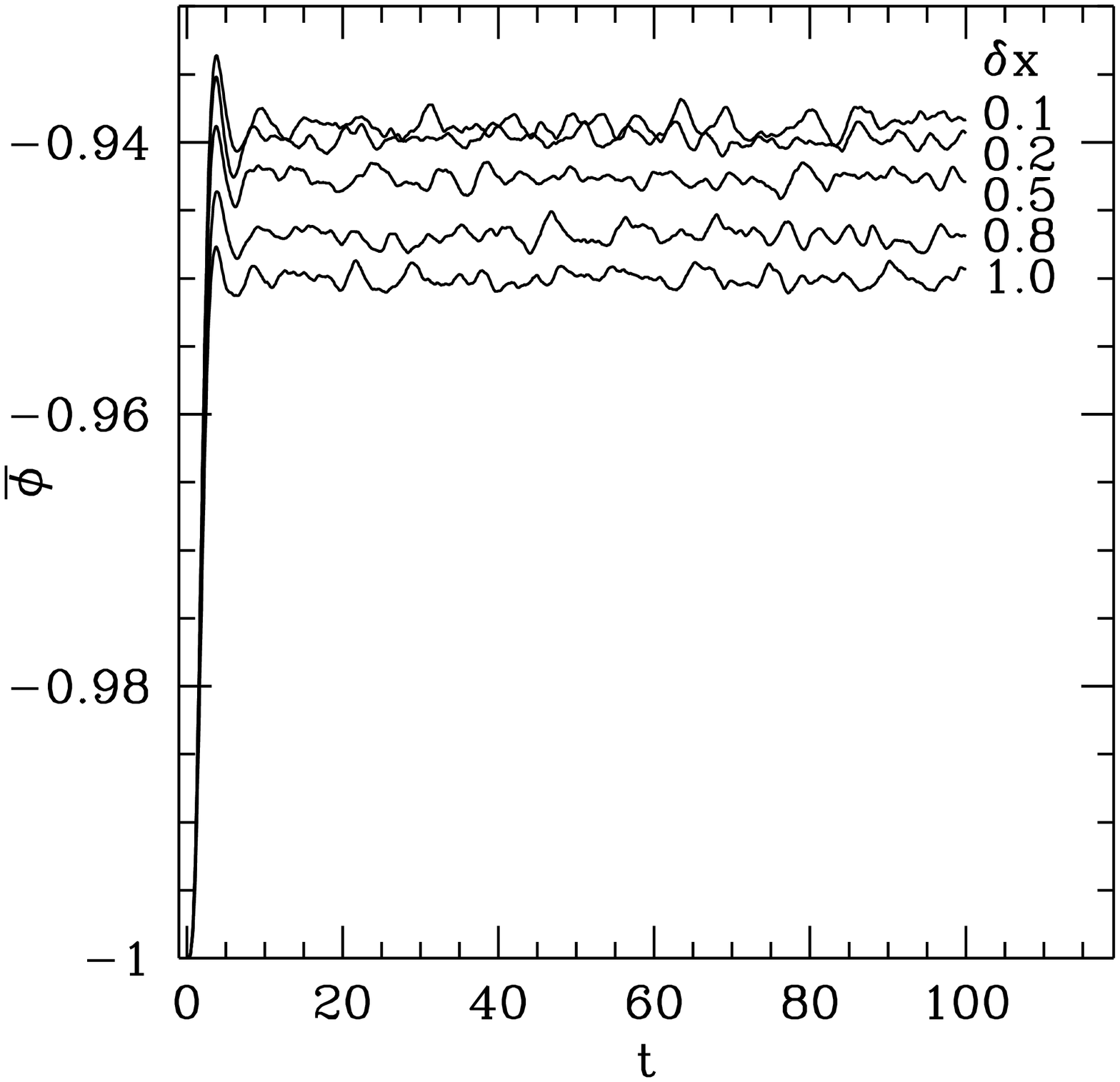,width=\linewidth}
\end{minipage}
\begin{minipage}[b]{.46\linewidth}
\psfig{figure=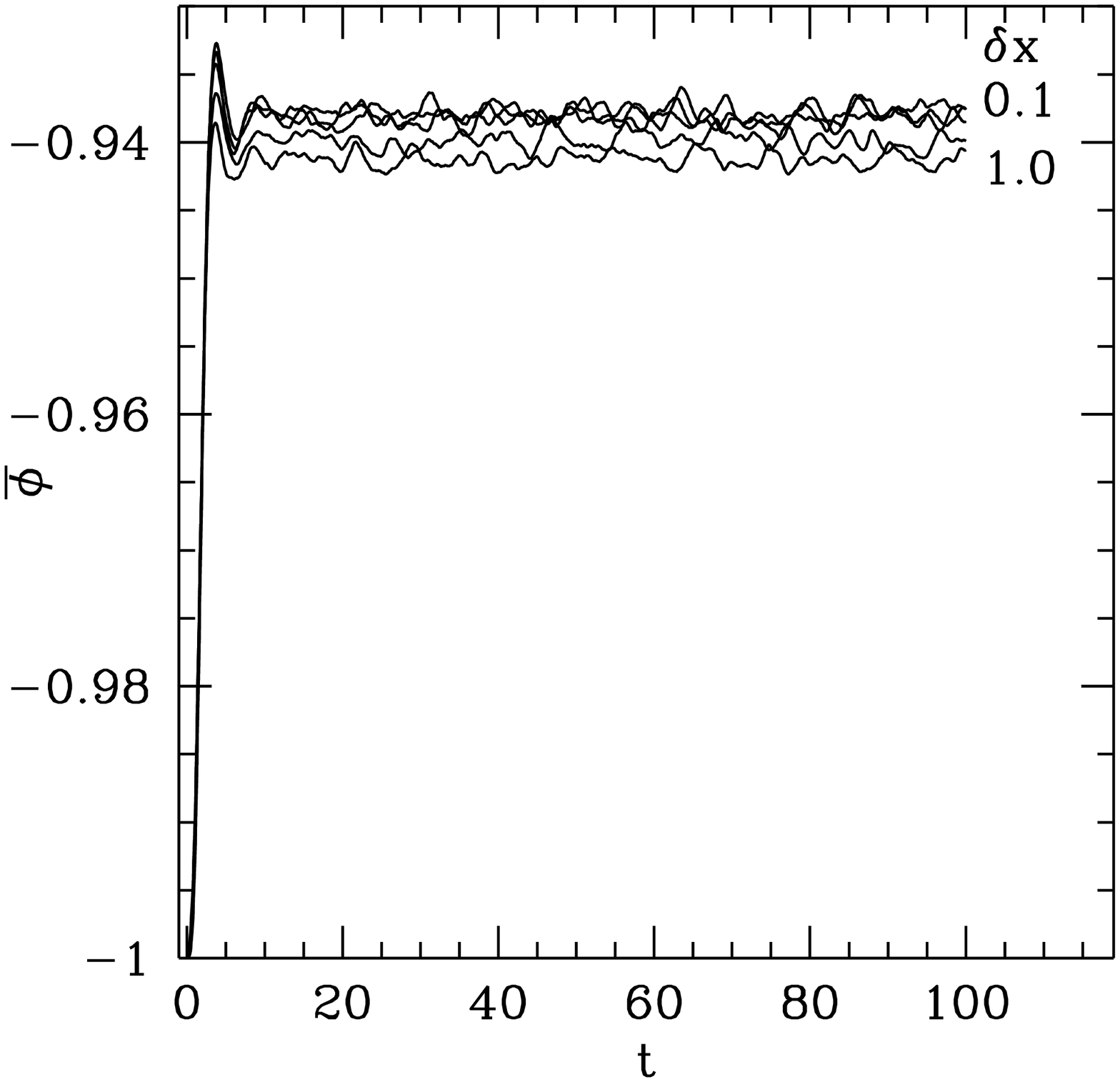,width=\linewidth}
\end{minipage}
\caption[]{Average field value $\bar\phi (t)$ for $T=0.1$ 
using the tree-level potential,
left, and the corrected potential, right. 
\label{f.phiave}}
\end{figure}

As a first test of our procedure, we investigate the mean field
value $\bar\phi (t)=(1/L) \int\phi (x,t) dx$ {\it before} the nucleation of a 
kink-antikink pair, {\it i.e.}, while the field is still well localized in 
the bottom of the well.
In Fig.\ \ref{f.phiave} we show the ensemble average of $\bar\phi$
(after 100 experiments) for different values of $\delta x$, ranging
from 1 down to 0.1, at $T = 0.1$.
The simulations leading to the left graphs use the ``bare'' potential
$V_0$, whereas the right graphs are produced employing $V_{\rm latt}$
(Eq.\ \ref{e.VLatt}).
Apart from a discrepancy for very coarse grids 
($\delta x = 1$), where the resolution nears the correlation length,
the average field
value is clearly lattice-spacing independent when using
$V_{\rm latt}$,
in contrast to the use of $V_0$.

As discussed before,
the average mean field value should correspond to the minimum
$-\phi_{\rm min}(T)$ of the effective potential. 
However, since we are only using a one-loop approximation, this agreement will
get progressively worse as the temperature increases. For example, for $T=0.2$,
the discrepancy between the theoretical value, $-\phi_{\rm min}(0.2)$, and the
numerical result is 10\%. For higher temperatures, we should not trust the
one-loop approximation; other nonperturbative effects, such as subcritical
fluctuations, too small in width and amplitude to emerge as a kink-antikink 
pair but
still large enough to bring the average 
value of the field away from its one-loop
value, will become important\cite{GLEISER-HECKLER}. Thus, we restrict our
investigation to temperatures safely within the 
limits of validity of the one-loop
approximation. In a subsequent study, we intend to investigate the role of
these nonperturbative effects.


{\bf Density of kink-antikink pairs.}
Perhaps the most difficult task when counting the number of kink-antikink
pairs that emerge during a simulation is the identification of what precisely is
a kink-antikink pair at
different temperatures. Typically, we can identify
three ``types'' of fluctuations: i) small amplitude, perturbative fluctuations
about one of the two minima of the potential; ii) full-blown kink-antikink
pairs interpolating between the two minima of the potential; 
iii) nonperturbative fluctuations which have large amplitude but not quite
large enough to
satisfy the boundary conditions required for a kink-antikink pair. These latter
fluctuations are usually dealt with by a smearing of the field over a certain
length scale. Basically, one chooses a given smearing length $\Delta L$ which
will be large enough to ``iron out'' these ``undesirable'' fluctuations but not
too large that actual kink-antikink pairs are also ironed-out. 
In this study, a similar smoothing was implemented as a four-pole
Butterworth low-pass  filter of the field with a filter cutoff length
$\Delta L$. The filter removes fluctuations with wavelengths smaller
than $\Delta L$. 
The choice of
$\Delta L$ is, in a sense, more an art than a science, given our ignorance
of how to handle these nonperturbative fluctuations. 

In Table 1 we show the
number of pairs for different choices of filter cutoff length and for different
temperatures. 
We counted pairs by identifying the zeros of the 
filtered field. From Table 1 it is clear that as the temperature
increases, the discrepancies in the count of pairs also increase. For
this reason we only trust our data for fairly low temperatures. The problem
is agravated by the fact that the ``size'' of the kink-antikink pair, 
{\it i.e.}, the minimal separation between the two, not only
changes due to dynamical effects, but also changes with temperature.
Thus, choosing the filter cutoff
length to be too large may actually undercount the number of pairs.
Choosing it too low may include nonpertubative fluctuations as pairs. We
chose $\Delta L=3$ in the present work, as this is the smallest ``size'' for
a kink-antikink pair. In contrast, in the works by Alexander et al. a
different method was adopted, that looked for zero-crossings for eight lattice
units (they used $\delta x=0.5$) to the left and right of a zero crossing 
\cite{HABIB}. 
We have checked that our simulations reproduce the results of Alexander et al.
if we: i) use the bare potential in the lattice simulations and ii) use a
large filter cutoo length $\Delta L$. Specifically, the number of pairs found
with the bare potential for $T=0.2,~\delta x=0.5$ are: $n_p = 36,~ 30,~{\rm
and}~ 27$, for $\Delta L = 3,~5,~{\rm and}~ 7$, respectively. Alexander et al.
found (for our lattice length) $n_p = 25$. Comparing these with Table 1,
it is clear that the differences between our results and those of Alexander
et al. come from using a different potential in the simulations, {\it viz.}
a corrected vs. an uncorrected potential.

We believe that at this point it is fair to say that the ``smearing issue''
remains unresolved, at least for temperatures $T>0.25$ or so. 
We intend to address the issue of how to deal with
these nonperturbative effects in a forthcoming
publication. In any case, the focus of the present work is mostly on how
to achieve a lattice-independent count, irrespective of the particular method
used for identifying the kink-antikink pairs.

\begin{figure} 
\begin{minipage}[b]{.46\linewidth}
\psfig{figure=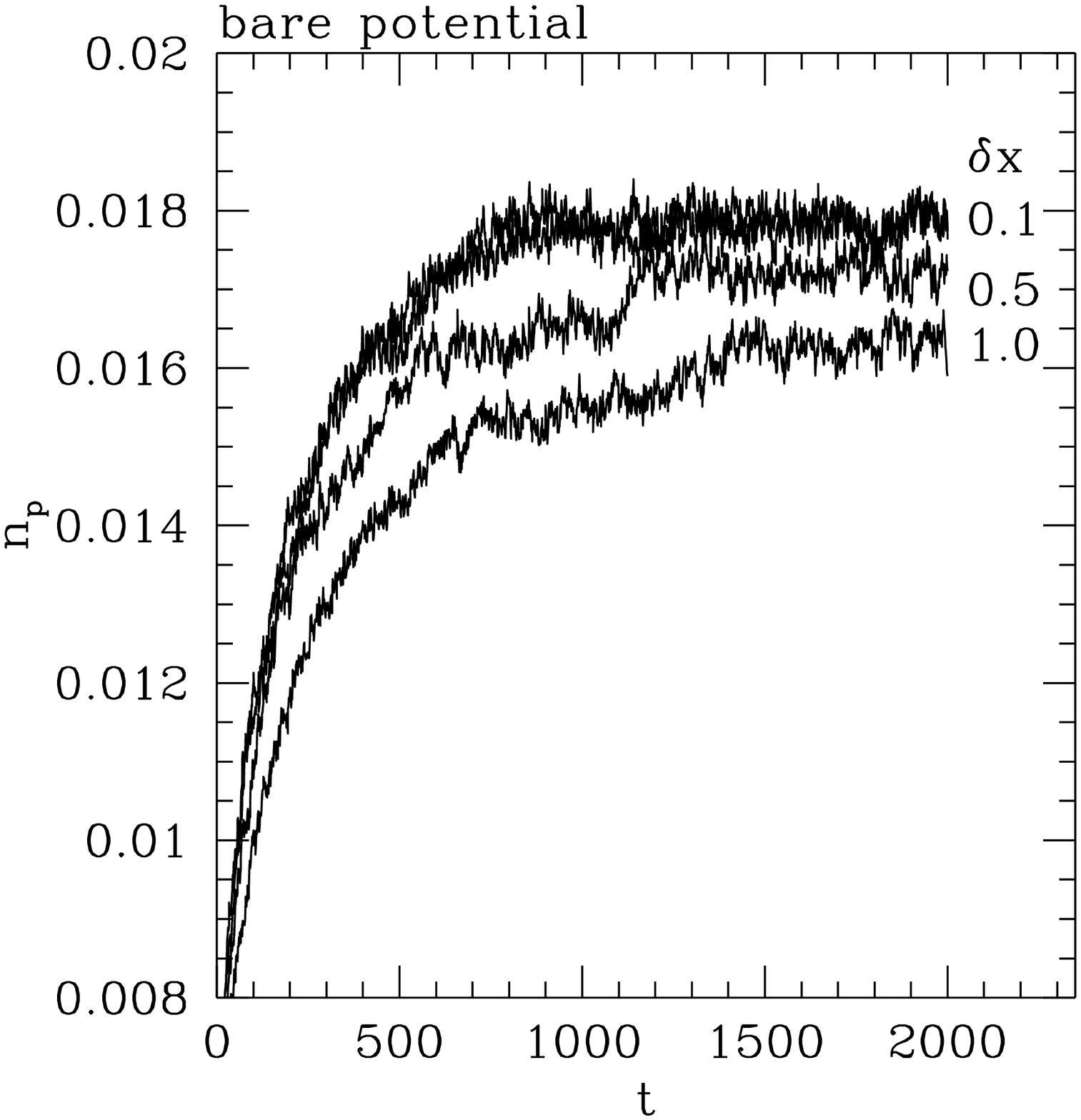,width=\linewidth}
\end{minipage}
\begin{minipage}[b]{.46\linewidth}
\psfig{figure=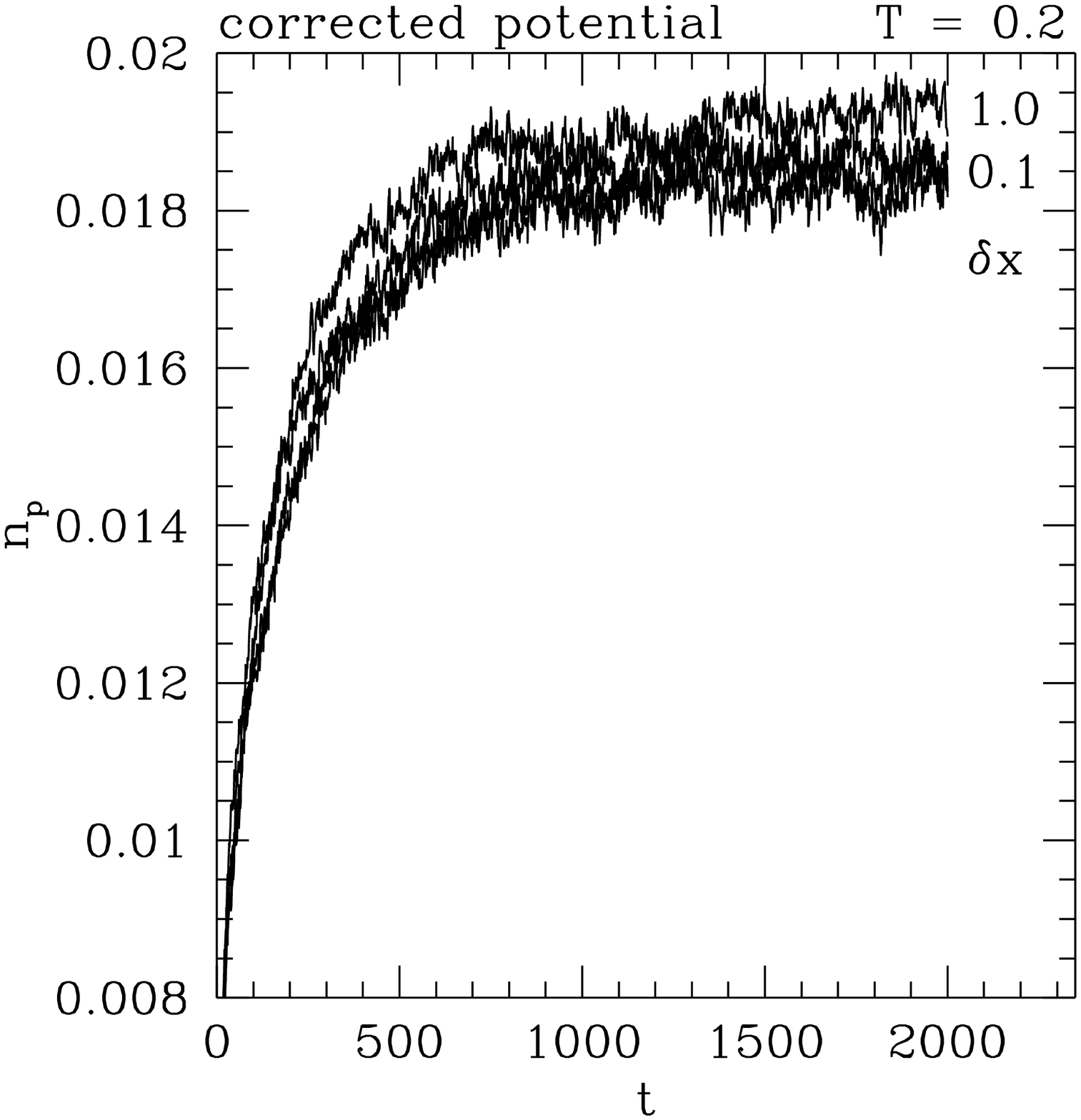,width=\linewidth}
\end{minipage}
\caption{\label{f.kinkdens}Density of kink-antikinks 
(half of density of zeros), for $T=0.2$
and $\delta x = 1$, 0.5, 0.2, and 0.1.}
\end{figure}

Fig.\ \ref{f.kinkdens} compares measurements of the kink-antikink 
pair density (half
the number of zeros of the filtered field), ensemble-averaged 
over 100 experiments,
for different lattice spacings.
Again it is clear from the graphs on the left
that using the tree-level potential 
$V_0$ in the
simulations causes the results to be dependent on $\delta x$,
whereas the addition of the finite counterterm 
removes this problem quite efficiently;
both diagrams of Fig.\ \ref{f.kinkdens} contain four
graphs each, 
although the graphs on the right are almost indistinguishable. Unless the
properly corrected potential is used in the lattice simulations, the measured
number density of topological defects is sensitive to the lattice
spacing. One must be careful when counting kinks, especially for large
lattice spacings, say $\delta x=0.25$ or larger.

The next step is to extract the correct continuum theory from the lattice
simulations. What theory is the lattice simulating? Most 
previous simulations of
thermal nucleation of kink-antikink pairs have
overlooked this
problem. Although a temperature-dependent kink mass was conjectured in
the works of reference \cite{BOCHKAREV}, not much has been done to
understand its origin or its value.
One way of addressing it is by comparing the numerically measured
kink mass with its theoretical prediction. 
It has been found that
the measured mass was smaller than the theoretical prediction by a factor
ranging 
from 25\% to 45\%\cite{BOCHKAREV,ALF-FELD-GLEI}, 
a disturbing result. This has been
attributed to 
several effects, such as the finite size of the lattice, the finite
size of the kinks, and phonon dressing effects due to the lattice
discretization\cite{MARCHESONI}. 
We will show that this problem is rooted in
the incorrect matching between theory and numerical simulations.
In the works by Alexander et al. a beautiful
agreement between the low temperature limit and a $T=0$ WKB approximation
was obtained, as well as between high temperatures and a double Gaussian
nonperturbative method \cite{HABIB}. Our method is effective precisely
between these two regimes, and could be interpreted as a $T$-dependent WKB
approximation obtained naturally from the inclusion of counterterms.

One should expect the equilibrium kink-antikink pair density to follow the
proportionality \cite{r.kinkdens}
\begin{equation}
n_{\rm kink} \propto {1\over \sqrt{T}} \exp (-M_k /T)
\end{equation}
where $M_k$ is the kink mass, given by
\begin{equation}
M_k = \int dx \left[ {1\over 2} {{\phi_k}'}^2 + V(\phi_k) \right] \, ,
\end{equation}
and $\phi_k(x)$ is the kink solution to the equation of motion. Note that we
left the potential $V(\phi_k)$ unspecified. If we use the tree-level
potential, $V_0(\phi_k)$, we obtain the well-known result
$M_k=\sqrt{8\lambda /9}\phi_0^3$. Or, in dimensionless variables,
$\tilde M_k=\sqrt{8/9}$. One can extract the numerical
value of $M_k$ by measuring the pair density and plotting the results in a
logarithmic scale, as in Ref. \cite{ALF-FELD-GLEI}. The result should be a
straight line with negative slope $-\tilde M_k=-\sqrt{8/9}$. 
However, as mentioned
above, the measured slope was found to be
about $-0.70\tilde M_k$. The reason for the
discrepancy is that the potential which should be used when
comparing theory and simulation is not the tree-level potential $V_0(\phi)$
but the effective potential $V_{\rm 1L}(\phi)$. 
Thus, one must compute the effective kink mass
$M_k(T)$ using the corrected potential
$V_{\rm 1L}(\phi)$ and {\it then}
compare the results with the numerical simulations.
 
\begin{figure}
\begin{center}
\begin{minipage}[b]{.94\linewidth}
\psfig{figure=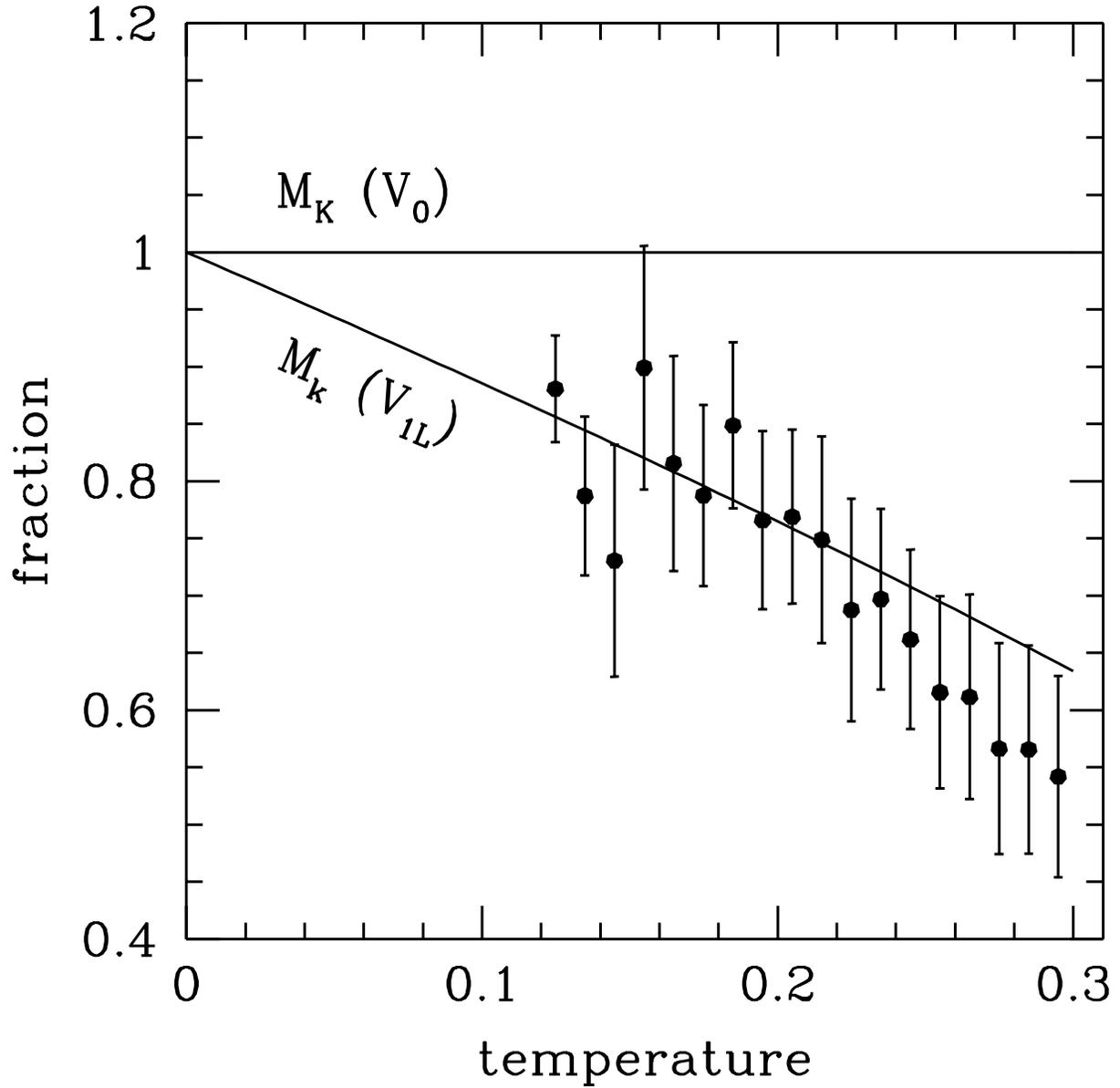,width=\linewidth}
\end{minipage}
\end{center}
\caption{\label{f.kinktheory}The ratio of the effective kink mass, $M_k(T)$, to 
the uncorrected kink mass, $M_k$, vs. the temperature.}
\end{figure}

The effective kink mass can be found
using the equation of motion and the real part
of $V_{\rm 1L}$ \cite{WEINBERG-WU},
\begin{equation}
M_k(T) = \int dx \left[ {1\over 2} {{\phi_k}'}^2 +
{\rm Re}(V_{\rm 1L}) \right] 
= 2\int_0^{\phi_{\rm min}}
\sqrt{2 {\rm Re}(V_{\rm 1L}(\phi))} \, d\phi \; .
\end{equation}
%
This integration can easily be carried out numerically. In
Fig.\ \ref{f.kinktheory} we plot the ratio $\tilde M_k(T)/\tilde M_k$ vs.\ the
dimensionless temperature, $\Theta$. Of course, for $T=0$, $\tilde M_k(0)/
\tilde M_k=1$. As the temperature increases, the effective kink mass
decreases. The points represent the kink mass extracted from the numerical
simulations, while the error bars were obtained by propagating the standard
deviation of the ensemble average. 
It is quite clear that the
effective kink mass tracks the numerical values quite well. In fact, within 
the validity of our approximations, the ``averaged'' value for the effective 
kink mass is $0.75 \tilde M_k$. Also, since the mass extracted from the
simulations depends on the filter cutoff
length $\Delta L$, the reasonable agreement 
between theory and
numerical experiment offers indirect support for our choice of $\Delta L=3$.
For very small and very large temperatures the theory fails to
track the numerical data. At large temperatures
$\Theta\geq 0.25$, the one-loop approximation breaks down, while for low
temperatures $\Theta\leq 0.12$, 
the large pair nucleation time-scale precludes a proper
statistical analysis (not enough experiments). 
However, the conclusion is quite clear: by controlling
the dependence on lattice spacing of the simulations we were able,
within the validity of our approximations, to obtain the correct effective
potential that should be used when comparing theory and numerical experiment.

MG was partially supported by the 
National Science Foundation through a
Presidential Faculty Fellows Award no. PHY-9453431 and by the 
National Aeronautics and Space Administration
grant no. NAGW-4270. HRM was supported by a
National Science Foundation
grant no. PHY-9453431 and by the National Aeronautics and Space
Administration grant no. NAGW-4270.

{\LARGE
\renewcommand{\arraystretch}{1.1}
\begin{tabular}{@{~}c@{~}|@{~}l@{~$\pm$~}l@{~}@{~}l@{~
$\pm$~}l@{~}@{~}l@{~$\pm$~}l@{~}}
$\Delta L$ & \multicolumn{2}{c}{$T = 0.15$} &
\multicolumn{2}{c}{$T = 0.20$} &
\multicolumn{2}{c}{$T = 0.25$} \\ \hline
3 & 10.3 & 0.2 & 39.0 & 0.5 & 75.6 & 0.6 \\
5 &  8.9 & 0.2 &  32.7  & 0.5 & 62.0 & 0.5\\
7 & 8.4 & 0.2 & 29.8  & 0.4 & 54.6 & 0.5\\
\end{tabular}}

\bigskip\bigskip\large
Table 1: Number of kink-antikink pairs for different choices of the filter
cutoff length $\Delta L$ and $T$.

\end{document}